\documentclass[
reprint,
superscriptaddress,
 amsmath,amssymb,
 aps,
prb,
floatfix,]{revtex4-2}
\usepackage{multirow}
\usepackage{graphicx}
\usepackage{dcolumn}
\usepackage{bm}
\usepackage{braket}
\usepackage[english]{babel}
\usepackage{xcolor}
\usepackage{bbold}
\usepackage{hyperref}
\usepackage{tikz}
\usepackage{color}

\begin{document}

\title{Magnetic order through Kondo coupling to quantum spin liquids}
\author{M. A. Keskiner}
\email{akif.keskiner@bilkent.edu.tr}
\affiliation{Department of Physics, Bilkent University, Ankara, 06800, T\"urkiye}
\author{M. \"{O}. Oktel}
\email{oktel@fen.bilkent.edu.tr}
\affiliation{Department of Physics, Bilkent University, Ankara, 06800, T\"urkiye}
\author{Natalia B. Perkins}
\email{nperkins@umn.edu}
\affiliation{School of Physics and Astronomy, University of Minnesota, Minneapolis, MN 55455, USA}
\author{Onur Erten}
\email{onur.erten@asu.edu}
\affiliation{Department of Physics, Arizona State University, Tempe, AZ 85287, USA}

\begin{abstract}
We study the emergence of magnetic order in localized spins that interact solely through their coupling to a Kitaev-type spin liquid.  Using three toy models 
-- the Kitaev model, the Yao-Lee model, and a square-lattice generalization of the Kitaev model
-- we calculate the effective exchange Hamiltonians mediated by the fractionalized excitations of these spin liquids. 
This setup is analogous to a Kondo lattice model, where conduction electrons are replaced by itinerant Majorana fermions. In the Kitaev model, our results show that the lowest-order perturbation theory generates short-range interactions with modified couplings and extending to sixth order introduces longer-range interactions while preserving the quantum spin-liquid ground state. 
Models involving more  Majorana flavors on honeycomb and square lattices exhibit more complex behavior. The honeycomb  Yao-Lee model with three flavors of itinerant Majorana fermions generates long-range RKKY-type interactions, leading to antiferromagnetic order and partial gapping of the Majorana fermion spectrum. In contrast, the square-lattice model produces a combination of anisotropic short- and long-range interactions, which can give rise to either a dimerized quantum paramagnetic state or an Ising antiferromagnet, depending on the parameters. These results illustrate the rich variety of magnetic orders that can be mediated by Kitaev-type spin liquids.
\end{abstract}

\maketitle

\section{Introduction}
Local moment magnetism in metallic systems is generally described by the Kondo lattice model where the interaction among the local moments is primarily mediated by the Ruderman–Kittel–Kasuya–Yosida (RKKY) interaction\cite{Ruderman_PR1954, Yosida_PR1957, Kasuya_PTP1956}. RKKY interaction arises due to second order processes where the local moments couple to the conduction electrons via the antiferromagnetic Kondo coupling. While the Kondo model has traditionally been applied to understanding the behavior of lanthanide and actinide intermetallics \cite{Coleman_book}, recently synthetic versions of Kondo lattices have been realized in van der Waals heterostructures including  1T/2H-TaS$_2$ \cite{Ayani_Small2024}, NbSe$_2$ \cite{Liu_SciAdv2021} as well as  1T/1H-TaSe$_2$ \cite{Wan_NatComm2023}, TaS$_2$  \cite{Vano_Nature2021} heterobilayers where 1T phases are Mott insulators and 1H and 2H phases are metals. Tunneling spectroscopy indicates the formation of Kondo resonances in these systems.  Synthetic Kondo lattices have also been proposed for graphene and two-dimensional magnet heterobilayers \cite{Leeb_PRL2021, Keskiner_NanoLett2024, Das_arXiv2024}.

\begin{figure}[t]
\includegraphics[width=0.9\linewidth]{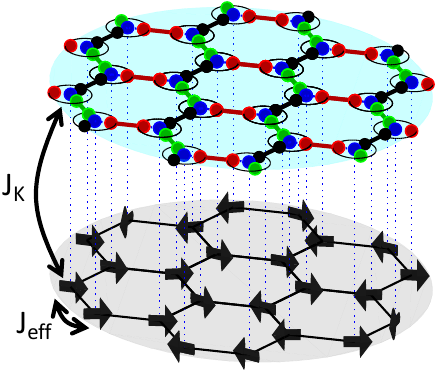}
    \caption{Schematic of our setup. Local moments in the bottom layer are coupled to a Kitaev-type spin liquid in the top layer via a Kondo coupling  ($J_K$). The effective exchange interactions between the local moments ($J_{\rm eff}$)   and resulting magnetic order are mediated by the fractionalized excitations of the spin liquid.}
    \label{Fig:1}
\end{figure}


 Motivated by these recent developments, we investigate the magnetic order in heterostructures composed of quantum spin liquids (QSLs) and Mott insulators. Our setup, illustrated in Fig.~\ref{Fig:1}, is conceptually similar to synthetic Kondo lattices but substitutes the metallic layer with a QSL layer.  There are promising candidates for QSLs among Kitaev materials such as $\alpha$-RuCl$_3$~\cite{HwanChun_NatPhys2015, Takagi_NatRevPhys2019,Kitagawa_Nat2018}. In addition, indications of single ion Kondo effect has been observed in Cr doped $\alpha$-RuCl$_3$ above the N\'eel temperature \cite{Lee_NatComm2023}. While the single impurity Kondo effect in Kitaev spin liquids has been explored extensively \cite{Dhochak_PRL2010, Vojta_PRL2016, Das_PRB2016}, 
the question of whether long-range magnetic order can be induced by coupling to Majorana fermions remains largely unexplored.

Here, we examine the mechanisms
by which non-local exchange interactions between localized spins and impurities are mediated by fractionalized excitations in QSLs. 
 These excitations, including spinons, Majorana fermions, and gauge bosons, differ fundamentally from conduction electrons, enabling unconventional long-range couplings.  
While the exchange couplings induced between the magnetic moments of the Mott insulator are expected to share similarities with traditional RKKY interactions, the mechanism of their formation might be different due to the nature of the fractionalized quasiparticles and the lack of conventional Fermi surfaces in some QSLs \cite{Legg2019,Zheng2021}. In these systems, the oscillatory and distance-dependent characteristics of the exchange couplings are dictated by the structure of the QSL's fractionalized excitation spectrum, rather than by a conventional Fermi surface, reflecting the unique physics of these quantum phases.

 In this work, we focus on the Kitaev-type spin liquids, which allow for an exact solution. Specifically, we are interested in the Kitaev model on the honeycomb lattice \cite{KITAEV20062},  the Yao-Lee (YL) model \cite{Yao_PRL2011}, and the generalization of the Kitaev model on a square lattice \cite{Nakai_PRB2012}. These models, sometimes referred to as 
 $\Gamma$-matrix generalizations of the Kitaev model \cite{Wu_PRB2009}, provide ideal platforms for investigating how fractionalized excitations in QSL induce exchange interactions among local moments coupled to the QSL via Kondo interaction. Our main results are the following: (i) A single flavor of Majorana fermions in the Kitaev model is unable to support long-range spin polarization. Therefore, the induced interaction is short-ranged and can be calculated via perturbation theory. The induced Hamiltonian is another Kitaev model with a QSL ground state. (ii) Among the $\Gamma$-matrix generalizations, YL model gives rise to long-ranged RKKY-type exchange. Our analysis indicates that this effective interaction is identical to the RKKY interaction in charge-neutral graphene up to prefactors. The ground state is a Heisenberg antiferromagnet (AFM) that gaps the excitation spectrum of two out of the three itinerant Majorana fermion flavors. (iii) The $\Gamma$-matrix generalization on the square lattice share similarities to both Kitaev and YL models as it mediates both short- and long-range interactions. The short-range interaction is only over one of the four types of bonds, and it is significantly stronger, which leads to dimerization and a quantum paramagnetic ground state. However, an Ising AFM ground state can be stabilized upon tuning the flux gap.
\begin{figure}[t]
\includegraphics[width=1\linewidth]{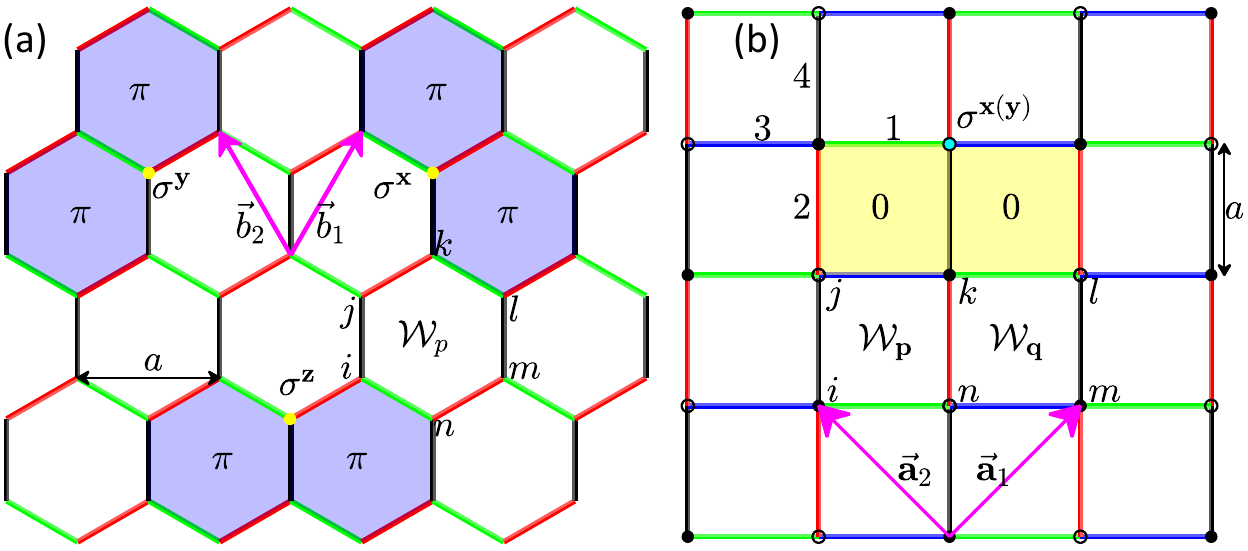}
    \caption{(a) The Kitaev and Yao-Lee models on the honeycomb lattice with three distinct bonds shown in  red, green and black. (b) The generalized Kitaev model on the square lattice with  four inequivalent bonds depicted  in red, green, blue and black.
    The primitive vectors for the honeycomb lattice are $\vec{b}_{1,2}=(\pm1/2,\sqrt{3}/2)a$, and for the square lattice, $\vec{a}_{1,2}=(\pm1,1)a$, where $a$ refers to lattice constant specific to each lattice. The application of the spin operators $\sigma^{\alpha}$  on the ground state of the Kitaev model flips the bond operators and creates two flux excitations (white and blue hexagonal plaquettes  represent $0$- and $\pi$-fluxes, respectively).
     Similarly in (b),  $\sigma^{x(y)}$ flips the black bonds and creates two visons that are $0$-flux excitations in this case (white and yellow square plaquettes  represent $\pi$- and $0$-fluxes, respectively). The bond-dependent exchange interactions in the square lattice effectively double the elementary unit cell, resulting in the presence of two inequivalent elementary plaquettes, labeled $p$ and $q$, as defined in the text.} 
    \label{Fig:2}
\end{figure}

\section{Microscopic Model and Method}
The setup, where a Kitaev-type quantum spin liquid is coupled to local moments at each site via Kondo coupling, is illustrated in Fig.~\ref{Fig:1}. The corresponding Hamiltonian is given by:
\begin{equation}
    H=H_{SL}+V
    \label{eq:H}
\end{equation}
where $H_{SL}$ represents the Hamiltonian of the QSL, and $V$ denotes the interaction between local moments and the spin liquid, $V=J_K\sum_j {\bf S}_j\cdot{\boldsymbol\sigma}_j$. Here, $J_K$ is the coupling strength, ${\bf S}_j$ and $\boldsymbol{\sigma}_j$ are Pauli matrices corresponding to the spin operators of the local moment and the spins in the QSL, respectively. For simplicity, we
assume the local moments to be non-interacting, as our
focus lies on the interactions mediated by the QSL. We consider the following three exactly solvable models:
\begin{eqnarray}
    H_K &=& \sum_{\langle ij \rangle_\alpha} K_\alpha \sigma_i^\alpha \sigma_j^\alpha \label{eq:K}\\
    H_{YL} &=& \sum_{\langle ij\rangle_{\alpha}} K_\alpha(\boldsymbol{\sigma}_i \cdot \boldsymbol{\sigma}_j)\otimes\tau_{i}^{\alpha}\tau_{j}^{\alpha} \label{eq:YL}\\
    H_{SqL} &=& \sum_{\langle ij\rangle_{\alpha}} K_\alpha(\sigma_{i}^{x}\sigma_{j}^{x}+\sigma_{i}^{y}\sigma_{j}^{y})\otimes\tau_{i}^{\alpha}\tau_{j}^{\alpha} \label{eq:SqL}
\end{eqnarray}
Here, ${\langle ij\rangle_{\alpha}}$ represents different types of bonds between nearest neighbor sites. $H_K$ is the Kitaev model on a honeycomb lattice\cite{KITAEV20062}. $H_{YL}$ refers to the Yao-Lee model\cite{Yao_PRL2011}, which is a $\Gamma$-matrix generalization of the Kitaev model that is defined on a honeycomb lattice with spin-orbital interactions. Similarly, $H_{SqL}$ is also a generalization of the Kitaev model but defined on a square lattice\cite{Nakai_PRB2012}. 
In Eq.~(\ref{eq:YL}) and (\ref{eq:SqL}), the $\sigma$ and $\tau$ are both independent Pauli matrices corresponding to the spin and orbital degrees of freedom. Both $H_{YL}$ and $H_{SqL}$ are referred as $\Gamma$-matrix generalizations because they can be expressed in terms of the $\Gamma$-matrices that satisfy the Clifford algebra, $\{\Gamma_{i}, \Gamma_j \} = 2\delta_{ij}$:
\begin{equation}
    H_{YL / SqL}=-\sum_{\langle ij\rangle_{\alpha}} K_\alpha (\Gamma_{i}^{\alpha}\Gamma_{ j}^{\alpha}+\sum_{\beta=c+1}^{5}\Gamma_{ i}^{\alpha\beta}\Gamma_{ j}^{\alpha\beta}), 
\end{equation}
where $c$ is the coordination number, $c=4$ for the SqL model and $c=3$ for the  YL model \cite{Wu_PRB2009, Yao_PRL2009, Carvalho_PRB2018, Chulliparambil_PRB2020, Seifert_PRL2020, Chulliparambil_PRB2021, Nica_npjQM2023, Vijayvargia_PRR2023, Akram_PRB2023, Keskiner_PRB2023, Majumder_PRB2024, Poliakov_PRB2024}. Eqs.~(\ref{eq:YL}) and (\ref{eq:SqL}) can be obtained via the relations $\Gamma^{\gamma}=-\sigma^{y}\otimes\tau^{\gamma}$, $\gamma=1,2,3$; $\Gamma^{4}=\sigma^{x}\otimes\mathbb{1}$; $\Gamma^5=-\sigma^{z}\otimes\mathbb{1}$ and $\Gamma^{\alpha\beta}=i[\Gamma^{\alpha},\Gamma^{\beta}]/2$ for $\alpha<\beta$. The $\Gamma$-matrix generalizations of the Kitaev model may be realized in strongly spin-orbit coupled systems with a four-dimensional local Hilbert space \cite{Churchill_arXiv2024}.

Before we calculate the magnetic exchanges mediated by the QSL layer, we first briefly discuss the exact solutions to the models described in Eqs.~(\ref{eq:K}), (\ref{eq:YL}) and (\ref{eq:SqL}).  A key feature of these models is the presence of conserved Wilson loop operators, $\mathcal{W}_p$ defined around each plaquette.  These operators are formed by taking the product of bond operators along the edges of a plaquette. Importantly, the Wilson loop operators commute with each other,  $[\mathcal{W}_p, \mathcal{W}_{p' }]=0$ and with the Hamiltonian, $[\mathcal{W}_p, H ]=0$. As a result, they act as conserved quantities, partitioning the Hilbert space into distinct sectors characterized by their eigenvalues, $\pm1$. This structure simplifies the solution by reducing the problem to one of free fermions or other effective degrees of freedom within a fixed sector of the conserved quantities.

The plaquette operators in the Kitaev and the YL models can be written as $\mathcal{W}^K_p=\sigma_i^x\sigma_j^y\sigma_k^z\sigma_l^x\sigma_m^y\sigma_n^z$, $\mathcal{W}_p^{YL}=\mathbb{1}\otimes\tau_i^x\tau_j^y\tau_k^z\tau_l^x\tau_m^y\tau_n^z$ (shown as $\mathcal{W}_p$ in Fig~\ref{Fig:2} (a)). For the square lattice model the bond-dependent exchange interactions double the unit cell resulting in two distinct elementary plaquettes, $\mathcal{W}_p^{SqL}=\sigma_k^z\sigma_n^z\otimes\tau_i^z\tau_j^x\tau_k^z\tau_n^x$, and $\mathcal{W}_q^{SqL}=\sigma_k^z\sigma_n^z\otimes\tau_k^x\tau_l^z\tau_m^x \tau_n^z$ (depicted in Fig.~\ref{Fig:2} (b)).

The Kitaev model and its $\Gamma$-matrix  generalizations can be exactly solved by introducing Majorana fermion representations with four or six Majorana fermions per site, respectively. The spin operators are expressed as 
$\sigma_i^\alpha = i b_i^\alpha c_i$ ($\alpha = x,y,z$) for the Kitaev model, and
 $\Gamma_i^{\alpha}=ib_i^{\alpha}c_i$ ($\alpha = 1,2,3,4,5$)
 for the $\Gamma$-matrix generalizations. The corresponding Hamiltonians in the Majorana fermion representation are: 

\begin{eqnarray}
H_K &=& \sum_{\langle ij\rangle_{\alpha}}-K_\alpha iu_{ij}^{\alpha}c_ic_j \\
H_{YL} &=& \sum_{\langle ij\rangle_{\alpha}}-K_\alpha iu_{ij}^{\alpha}(c^x_ic^x_j+c^y_ic^y_j+c^z_ic^z_j) \label{eq:MFYL}\\
H_{SqL} &=& \sum_{\langle ij\rangle_{\alpha}}-K_\alpha iu_{ij}^{\alpha}(c^x_ic^x_j+c^y_ic^y_j) \label{eq:MFSqL}
\end{eqnarray}
where $u_{ij}^\alpha=ib_{i}^{\alpha}b_{j}^{\alpha}$. The quantity $u_{ij}^\alpha$ also commutes with the Hamiltonian, $[H, u_{ij}^\alpha] = 0$, making it a constant of motion. Also note that, to simplify the representation,  we relabel $b_{j}^{4}\rightarrow c_j^z$ in Eq.~(\ref{eq:MFYL}) as well as $c_{j}\rightarrow c_j^y$, and $b_{j}^{5}\rightarrow c_j^x$ in Eqs.(\ref{eq:MFYL}) and (\ref{eq:MFSqL}) \cite{Nica_npjQM2023, Vijayvargia_PRR2023}. 

The Majorana representation introduces an  overcomplete Hilbert space, and the physical 
states must satisfy the constraint $D_{i}=ib^x_{i}b^y_{i}b^z_{i}c_{i}=1$ for the Kitaev model and $D_{i}=ib^1_{i}b^2_{i}b^3_{i}b^4_{i}b^5_{i}c_{i}=1$ for the $\Gamma$-matrix generalizations. The SqL and YL models are described by two and three flavors of itinerant Majorana fermions, whereas the original Kitaev model has a single flavor. 

The plaquette operators can also be expressed in terms of the bond operators as $\mathcal{W}_p=(-i)^n\prod_{(j,k)\in p}u_{jk}^\alpha$ where $n$ is the number of the links on the boundary of the plaquette $p$ and the product is taken in the counter-clockwise direction  and their eigenvalues are $\pm1$. According to Lieb's theorem\cite{Lieb1994}, the ground states of Kitaev and YL model on honeycomb lattice lie in $0$-flux sectors ($\mathcal{W}_p = +1$) whereas the ground state of the model on square lattice ($H_{SqL}$) lies in $\pi$-flux sector ($\mathcal{W}_{p/q} = -1$). 

The commutation relations of the plaquette operators and the spin operators play a crucial role in determining the induced interactions between magnetic impurities. For the Kitaev model, the spin operators at site $i$, ${\sigma}_i^\alpha$, ($\alpha=x, y, z$), anticommute with the bond operator $u_{ij}^\alpha$, $\{\sigma_i^\alpha, u_{ij}^\alpha\}=0$. Therefore, the application of $\sigma_i^\alpha$ on the ground state of the Kitaev model flips $u_{ij}^\alpha$ and creates two vison excitations as depicted in Fig.~\ref{Fig:2}(a).

In contrast, in the YL model, Eq.(\ref{eq:YL}), the plaquette operators commute with each component of the spin operator,  $[\sigma_i^\alpha, \mathcal{W}_p]=0$, so that the Kondo coupling to the magnetic impurity does not generate flux in the YL QSL, allowing the system to remain in the flux-free ground state. Consequently, the microscopic processes mediating the RKKY interactions between local moments  coupled to  the YL model do not involve flux creation. This behavior  differs from the Kitaev model,  where these processes involve flux creation, making the mechanisms of interaction in the two systems qualitatively distinct.

In the SqL model, Eq.(\ref{eq:SqL}), operators $\sigma_i^z$ commute with all plaquette operators. However,  $\sigma_i^x$  and $\sigma_i^y$  anticommute with the bond operator $u_{ij}^4=ic_i^zc_j^z$, satisfying $\{ic_i^zc_i^y,ic_i^zc_j^z\}=0$ and $\{-ic_i^zc_i^x,ic_i^zc_j^z\}=0$. Consequently, applying $\sigma_i^x$  and $\sigma_i^y$ on the ground state flips the flux in the plaquettes that share the black bond where $u_{ij}^4$ is defined, as shown in Fig.~\ref{Fig:2}(b).

Building on this understanding, we calculate the effective exchange couplings, $J_{\rm eff}$, between  two local magnetic moments by treating the Kondo coupling $V$ in Eq.~(\ref{eq:H}) as a perturbation. For simplicity and consistency, we focus on the isotropic limit, where $K_\alpha = K$ for all bond directions.  
This choice not only simplifies the calculations but also  ensures the presence of gapless excitations, which are crucial for mediating longer-range interactions between the local moments. 
Nevertheless,  our results can be readily generalized to anisotropic cases with bond-dependent interactions, extending their relevance and applicability.

To compute $J_{\rm eff}$, we employ two complementary approaches based on the  system’s flux behavior. When the
spin operator generates gapped vison excitations, we apply perturbation theory to derive $J_{\rm eff}$. In the absence
of flux creation, we use exact diagonalization in the Majorana fermion representation, enabling calculations on relatively large systems.  For benchmarking, we also perform full  exact diagonalization of the spin Hamiltonian on small clusters. This combined approach provides a robust framework for systematically calculating the induced interactions between local moments mediated by the fractionalized excitations of the QSL layer. 


In the YL model, where spin operators commute with plaquette operators, we determine the induced interaction between impurity spins by calculating the energy difference between aligned and anti-aligned spin configurations using exact diagonalization of the Majorana fermion Hamiltonian, $J_{\rm eff}=(E_{\rm FM}-E_{\rm AFM})/2$. Our method is similar to RKKY calculations metallic systems \cite{ABS2010}. A similar approach is applied to magnetic impurities aligned in the $z$-direction on the square lattice model, where the $\sigma^z$ also commutes with the flux operator. 

For the Kitaev model, where the spin operator generates fluxes, we calculate $J_{\rm eff}$ perturbatively as follows:
  \begin{equation}
      \Sigma(E)=P_0(V+VG^{'}(E)V+VG^{'}(E)VG^{'}(E)V+...)P_0
  \end{equation}
where $P_0$ is the projection operator for the ground state and $G^{'} (E)=((E-H_0)^{-1})^{'}$ is the unperturbed Green's function for the excited states of $H_0$. The prime symbol $'$ indicates that the operator $((E-H_0)^{-1})^{'}$ acts only on the excited states while it is zero for the ground states. By setting $E=E_{GS}$, we calculate the first-order term, $H_{\rm eff}^{(1)}=P_0VP_0$, the second-order, $H_{\rm eff}^{(2)}=P_0VG^{'}(E)VP_0$, and higher-order corrections. 

In the square lattice (SqL) model, where the $x$ and $y$-components of the spin operator create flux excitations, we apply the same perturbative framework to compute the effective interactions.

\section{Results and Discussion}
\subsection{Magnetic impurities coupled to Kitaev model}

We begin by deriving induced interaction between magnetic impurities coupled to the Kitaev QSL. 
As discussed in the previous section, applying $\sigma_i^\alpha$ to the ground state of the Kitaev model flips the bond variable $u^\alpha_{ij}$ associated with that direction and creates two flux excitations  with an energy cost of $2\Delta \approx 0.26 |K|$\cite{KITAEV20062}. To restore the system to the flux-free sector, the lowest order process is to apply $\sigma_j^\alpha$ to the nearest neighbor site, which flips $u^\alpha_{ij}$ back to the ground state as shown in Appendix~\ref{App:A}, Fig.~\ref{Fig:8}(a). This leads to an induced exchange interaction,
\begin{eqnarray}
    H_{\rm eff}^{(2)}=\frac{J_K^{2}}{\Delta}
    \langle \sigma_i^{\alpha}\sigma_j^{\alpha} \rangle {S}_i^\alpha{S}_j^\alpha
\end{eqnarray}
The nearest-neighbor spin correlation, $\langle \sigma_i^{\alpha} \sigma_j^{\alpha} \rangle$, can be calculated exactly in the thermodynamic limit using Majorana fermion techniques. Its value is approximately $\langle \sigma_i^{\alpha} \sigma_j^{\alpha} \rangle \simeq {\rm sgn}(K) \cdot 0.525$, where ${\rm sgn}(K)$ denotes the sign function\cite{Baskaran2007}. A similar result for the effective exchange was obtained in Ref.~\citenum{Dhochak_PRL2010}. Importantly, since $\langle \sigma_i^\alpha \sigma_j^\alpha \rangle = 0$ for sites $i$ and $j$ that are not nearest neighbors, long-range interactions among the local moments are entirely absent.

\begin{figure}[t]
    \includegraphics[width=1\linewidth]{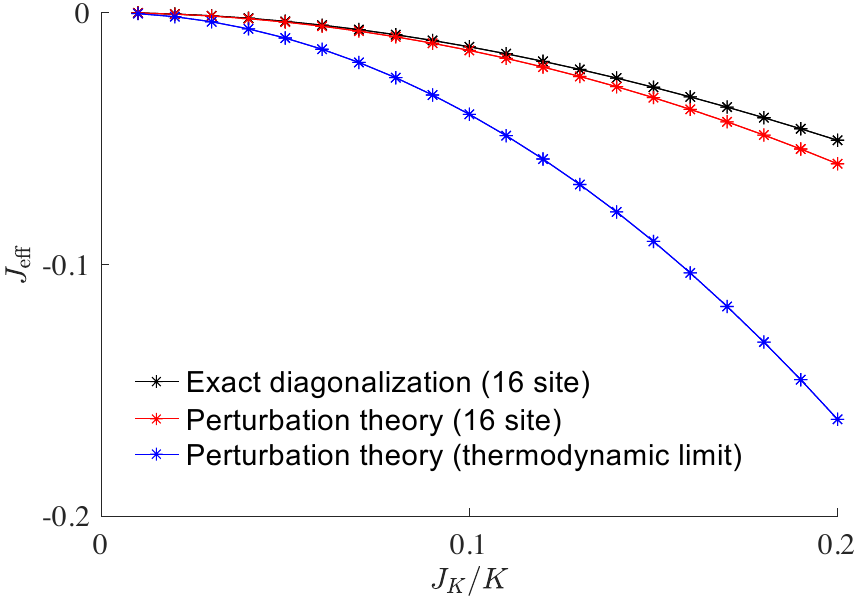}
    \caption{Comparison between the perturbation theory and exact diagonalization calculations for the effective interaction mediated by a Kitaev QSL. Exact diagonalization results for 16-site clusters show good agreement with perturbative estimates for 
    $J_K<\Delta$.  However, perturbative results in the thermodynamic limit deviate significantly, emphasizing the importance of finite-size effects.} 
    \label{Fig:3}
\end{figure}

To benchmark our perturbative results, we study a 16-site Kitaev model coupled to two nearest-neighbor local moments. Using exact diagonalization in the spin basis, we calculate the ground state energy as a function of $J_K/|K|$, considering the local moments aligned parallel ($E_{\rm FM}$) and antiparallel ($E_{\rm AFM}$).
We estimate the effective interaction strength as $J_{\rm eff}= (E_{\rm FM}-E_{\rm AFM})/2$. As shown in Fig.~\ref{Fig:3}, our perturbative estimation of  $J_{\rm eff}$, obtained  by 
 computing the nearest-neighbor spin-spin correlation and vison gap energy in the 16-site system, 
is quite close to the exact diagonalization result for $J_K/\Delta<1$. Note that the perturbative estimation for the thermodynamic limit differs significantly from the 16-sites systems, highlighting the importance of finite size effects.
 
Since the lowest order induced Hamiltonian, $H_{\rm eff}^{(2)}$, is simply another Kitaev model but higher-order terms are, in principle, nonzero, we examine these terms in perturbation theory to check the stability of the obtained spin liquid. At higher orders, certain processes return the flux sector to the ground state, and the corresponding induced Hamiltonian terms are derived in Appendix~\ref{App:A}. Due to time-reversal symmetry, all odd-order terms in the perturbation expansion vanish. Considering terms up to sixth order, the induced Hamiltonian is given by
  $H_{\rm eff} = H^2_{\rm eff}+H^4_{\rm eff}+H^6_{\rm eff}$ is given by
\begin{eqnarray}
  H_{\rm eff}&=&J_{\rm eff}^{(2)}\sum_{\langle ij\rangle_{\alpha}}  S_{i}^{\alpha}S{ j}^{\alpha} +\sum_{\substack{\langle ijkl\rangle \in p,\\{\alpha\neq\beta}}} J_{\rm eff}^{(4)} S_{i}^{\alpha}S_{ j}^{\beta}S_{k}^{\alpha}S_{ l}^{\beta}\nonumber \\
  &+&\sum_p J_{\rm eff}^{(6)}\mathcal{W}_p \label{eq:effKitaev}
\end{eqnarray}
where $J_{\rm eff}^{(2)} = \frac{J_K^2}{\Delta}\langle \sigma_i^{\alpha}\sigma_j^{\alpha} \rangle$, $J_{\rm eff}^{(4)} = \frac{3J_K^4}{\Delta^3}\langle \sigma_{i}^{\alpha}\sigma_{ j}^{\beta}\sigma_{k}^{\alpha}\sigma_{ l}^{\beta}\rangle$ and $J_{\rm eff}^{(6)} = 45\frac{J_K^6}{2\Delta^5}$, and $\langle ijkl\rangle \in p$ indicates the sum runs over all possible sets of four consecutive sites on a plaquette $p$, in order to preserve the ground state flux-free sector. The signs of $J_{\rm eff}^{(2)}$ and $J_{\rm eff}^{(4)}$ are dictated by the two- and four-spin correlations in $J_{\rm eff}^{(2)}$ and $J_{\rm eff}^{(4)}$, which, in turn, depend on the sign of $K$. Consequently, a ferromagnetic (FM) Kitaev model induces FM Kitaev interactions among the
local moments, while an antiferromagnetic (AFM) Ki-
taev model induces AFM Kitaev interactions.

Expressing the Hamiltonian in terms of Majorana fermions, the $J_{\rm eff}^{(4)}$ term leads to the third nearest neighbor hopping term for the itinerant Majorana fermions, but maintains the exact solvability of the model. By fixing the gauge to $u_{ij}=1$ for all bonds and performing a Fourier transform, the Hamiltonian becomes: 

\begin{equation}
    H=\sum_{k \in HBZ}\begin{pmatrix}
       a_{k}^{\dagger}&b_{k}^{\dagger} 
    \end{pmatrix}
    \begin{pmatrix}
        0 & i(f_k+g_k)\\
        -i(f_k^{*}+g_k^{*}) & 0
    \end{pmatrix}\begin{pmatrix}
        a_{k}\\b_{k}
    \end{pmatrix}
    \end{equation}
where the summation is over the half Brillouin zone, and $a_{k}$ and $b_{k}$ denote  fermions on the even and odd sublattices, respectively; $f_k=2J_2(1+e^{-ik_1}+e^{-ik_2})$ and $g_k=2J_4(e^{-i(k_1-k_2)}+e^{-i(k_1+k_2)}+e^{-i(k_2-k_1)})$ with $k_1={\bf k} \cdot {\bf b}_1$ and $k_1={\bf k}\cdot{\bf b}_2$ where ${\bf b}_1$ and ${\bf b}_2$ are the primitive lattice vectors (see Fig~\ref{Fig:2}(a)). The spectrum is given by $\epsilon(k)=\pm|f_k+g_k|$. Note that 
$J_{\rm eff}^{(4)}$ term does not open a gap in the spectrum of  $H_K$. Instead of gapping out the Majorana excitations, this term preserves the gapless nature of the system, leaving the Dirac points at their original locations in momentum space. The main effect of  $J_{\rm eff}^{(4)}$ is a slight renormalization of the Fermi velocity, modifying the dispersion of the low-energy excitations without altering the overall topological structure of the spectrum.

Our model of the Kitaev QSL coupled to spin-1/2 moments with induced Kitaev interactions resembles a bilayer Kitaev model but with two layers with very different values of the Kitaev coupling. Recent studies on the bilayer Kitaev model \cite{Seifert_PRb2018, Tomishige_PRB2019, Vijayvargia_PRB2024} suggest that the Kitaev spin liquid on each layer is stable up to a critical value of the interlayer exchange. Beyond this critical value, a gapped QSL \cite{Seifert_PRb2018} or a trivial interlayer singlet phase \cite{Tomishige_PRB2019} may arise.  Whether this conclusion holds in our scenario, with different Kitaev couplings on each layer, requires further investigation.

\begin{figure}[t]
    \includegraphics[width=1\linewidth]{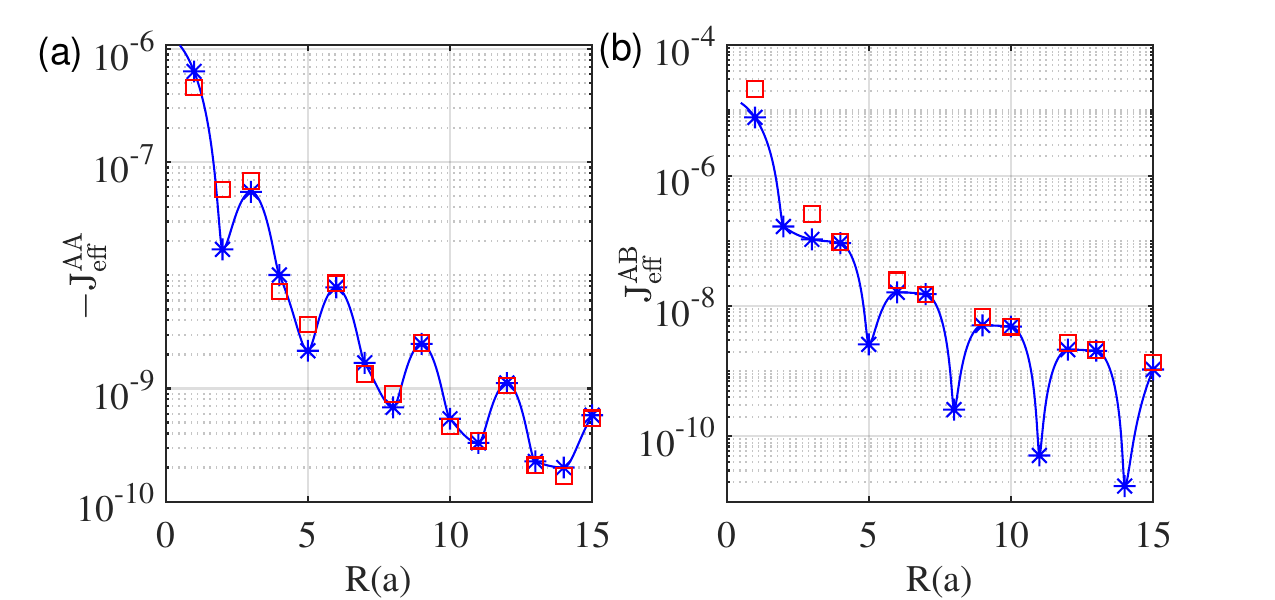}
    \caption{Induced long-range interactions Jeff among local moments in the Yao-Lee model are shown as a function of the
distance between them for (a) moments on the same sublattice (AA) and (b) moments on different sublattices (AB). The blue
stars are exact diagonalization results in Majorana fermion representation and the red squares are fits to Eq.(~\ref{eq:RKKYGraphene}).} 
    \label{Fig:4}
\end{figure}

\subsection{Magnetic impurities coupled to the Yao-Lee model} 

 We now calculate the induced interaction between magnetic impurities coupled to the YL QSL (\ref{eq:YL}). In the YL model, the spin operators and the flux operators commute. This property simplifies the problem because the flux sector  can be treated as static when analyzing spin dynamics. Namely, the Majorana fermions become the primary carriers of spin polarization in the system. These fermions mediate interactions between the magnetic impurities, similar to how conduction electrons mediate interactions in conventional metals through the RKKY mechanism.
In particular,  the induced interactions
resemble RKKY coupling in graphene at charge neutrality due to the similar excitation spectrum. 

 We compute the interactions between magnetic moments  by performing
large-scale exact diagonalization calculations using Majorana fermions. The results, shown in Fig.~\ref{Fig:4}(a) and (b),
correspond to interactions between moments on the same sublattice, $J_{\rm eff}^{AA}$, and on different sublattices,
$J_{\rm eff}^{AB}$.
  We show that the induced interaction decays  as  $1/|{\bf R}|^{3}$,
  where $|{\bf R}|$ is the separation between the moments, 
  and exhibits oscillations of the form $(1+\cos(2 {\bf k_D} \cdot \bf{R}))$, with  ${\bf k_D}=(\pm\frac{4\pi}{3a},0)$ representing the Dirac  momentum. This behavior can be well described by the following phenomenological form, originally derived for graphene at charge neutrality~\cite{ABS2010}:
\begin{eqnarray}
    J_{\rm eff}^{AA}(\bf{R})&=&-C\frac{J_K^2}{K}\frac{1+\cos(2\bf{k_D} \cdot \bf{R})}{|{\bf R}|^3} \label{eq:RKKYGraphene} \\ \notag
    J_{\rm eff}^{AB}(\bf{R})&=&C\frac{3J_K^2}{K}\frac{1+\cos(2\bf{k_D} \cdot \bf{R}+\pi)}{|{\bf R}|^3} 
\end{eqnarray}
where $C$ is a proportionality constant. The local moment, depending on its orientation, couples to only two flavors of Majorana fermions while the third flavor stays decoupled. Forming one complex fermion out of the two Majorana fermions maps our model to a graphene-like model where the spin up and down bands of graphene correspond to electron and hole bands of the new model (see Appendix~\ref{App:C}). This justifies the use of  Eq.~(\ref{eq:RKKYGraphene}). Notably, our prefactor, $C=0.0092$, obtained from the fit to the numerical results, is about four times larger than what is reported for graphene ($C=1/(72\sqrt{3}\pi)$), since we use Pauli matrices instead of $S=1/2$ operators.
The induced Hamiltonian is then given by:
\begin{eqnarray}
    H_{\rm eff} = \sum_{ij} J_{\rm eff}({\bf R}_i-{\bf R}_j) {\bf S}_i \cdot {\bf S}_j 
\end{eqnarray}
The SU(2) symmetry in $H_{\rm eff}$ energes from the inherent SU(2) symmetry of the YL model and Kondo coupling. As shown in Fig.~\ref{Fig:4}(a) and (b), $J_{\rm eff}$ is always antiferromagnetic for different sublattices and ferromagnetic for the same sublattice. Therefore, the ground state of the induced Hamiltonian is a Heisenberg antiferromagnet, which has long-range order at T=0. We note that although the exchange is long-ranged, its the strength is approximately two orders of magnitude weaker than the short-range interaction in Kitaev model.

We can also consider the back action of the magnetically ordered local moments on the Yao-Lee spin liquid.
The antiferromagnetic alignment of the local moments influences the Majorana fermion spectrum of the YL layer as it leads to a staggered field. For instance, if the local moments order along the $z$ axis, this effective field takes the form, $\sum_{j}J_{K}(-1)^{n_j}{\sigma_j^z}$ where $n_j = \pm 1$ for A/B sublattice. Since $\sigma_i^z = i c_i^x c_i^y$, it couples the $c^x$ and $c^y$ Majorana fermions and partially gaps the spectrum while the $c^z$ dispersion remains gapless. We note that the ordering along the $z$ axis is not crucial. Regardless of the axis along which the local moments align antiferromagnetically, two out of three Majorana fermion flavors become gapped since the Majorana fermions of the YL model exhibit O(3) symmetry.

\subsection{Magnetic impurities coupled to the square lattice (SqL) model}
\subsubsection{ Derivation of the couplings}
The SqL model as given in Eq.~(\ref{eq:SqL}) has similarities to both the Kitaev and the YL model. As in the Kitaev model, $\sigma^x$ and $\sigma^y$ operators anticommute with the flux operators,  resulting in short-range interactions. Yet, $\sigma^z$ commutes with $\mathcal{W}_p$,  leading to a long-range $z$-component of the induced interaction, similar to the behavior observed in  the YL model.

\begin{figure}[t]
\includegraphics[width=0.9\linewidth]{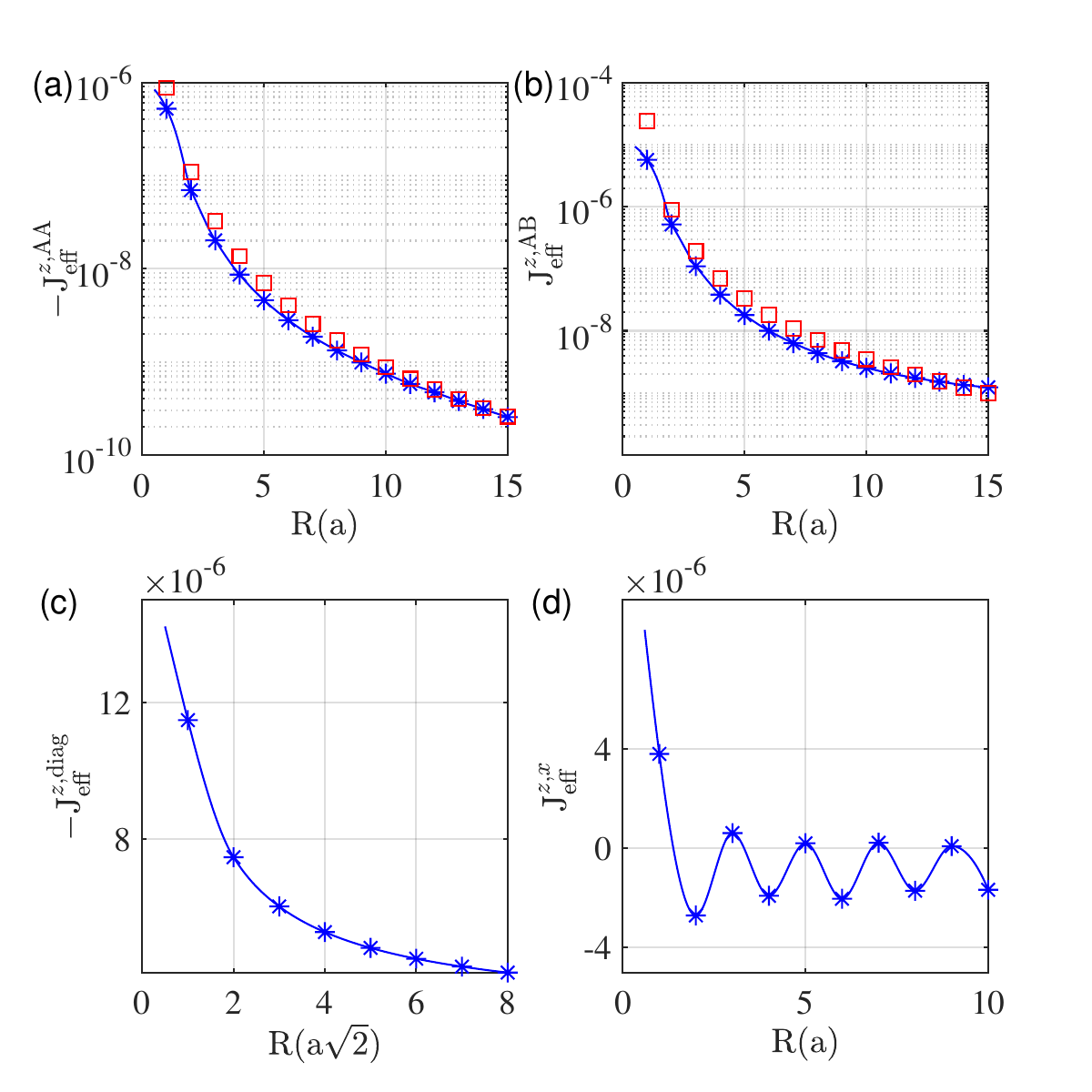}
    \caption{
    Induced long-range interactions $J_{\rm eff}$ (blue stars) between local moments in the square lattice (SqL) model for both $\pi$-flux and $0$-flux cases. (a) and (b) correspond to the $\pi$-flux square lattice, where (a) shows interactions between moments on the same sublattice (AA) and (b) between moments on different sublattices (AB). The red squares represent a numerical fit using the Dirac momentum of the $\pi$-flux dispersion in Eq.~\ref{eq:RKKYGraphene}.  (c) and (d) illustrate the induced long-range interactions in the $0$-flux square lattice, with (c) depicting ${\rm J_{eff}}^{z, {\rm diag}}$ along the diagonal direction and (d) showing ${\rm J}_{\rm eff}^{z,x}$ along the $x$-direction.} 
    \label{Fig:5}
\end{figure}
\begin{figure}[t]
\includegraphics[width=0.9\linewidth]{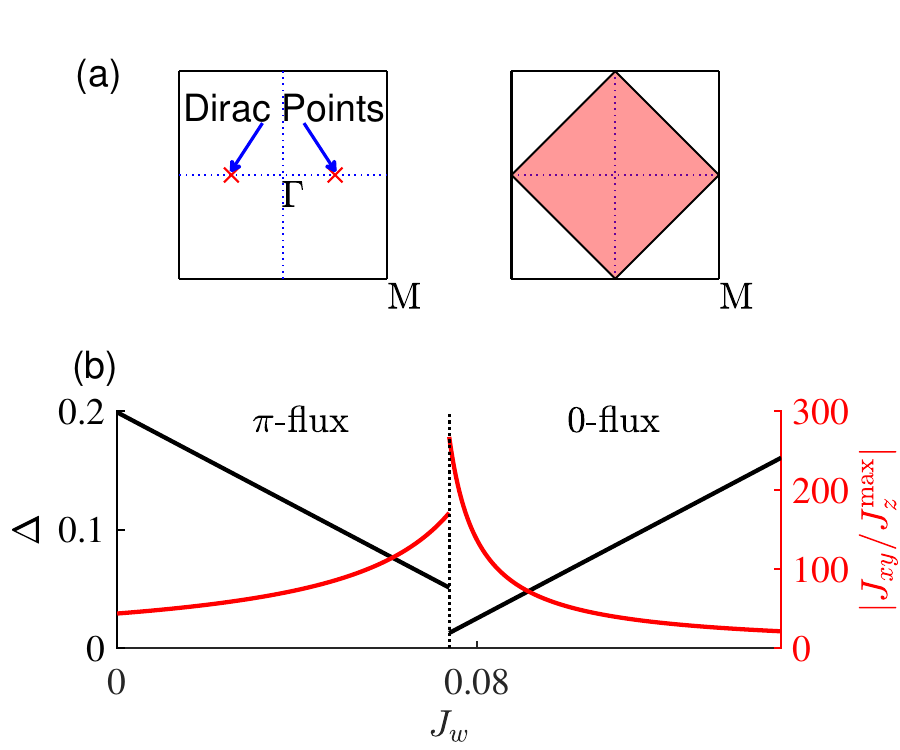}
    \caption{ (a) The $\pi$-flux square lattice features Dirac points, whereas the 0-flux square lattice exhibits a Fermi surface.
    (b) Dependence of the vison gap  energy $\Delta $ and the ratio 
     $J_{xy}/J_z^{ {\rm max}}$  on $J_w/|K|$. At $J_w=0.0738|K|$, we observe the change in the ground state from 
     $\pi$ to $0$-flux. Both $\Delta$ and $J_w$ are measure in the units of $|K|$. 
 }
    \label{Fig:6}
\end{figure}

We start our analysis with the short-ranged interactions. Once the magnetic impurities are ordered along either the $x$ and  $y$-directions, they produce effective magnetic field (proportional to  $J_K$) in the corresponding $x$ and  $y$-directions. The application of these effective fields generates fluxes on the plaquettes separated by black bonds (as depicted in Fig.~\ref{Fig:8}(f)). This occurs because the  $\sigma^x$ and $\sigma^y$ operators anticommute with the bond operators defined on these black bonds.
The second-order perturbation theory reveals that the induced interaction between the impurities linked by a black bond is given by 

\begin{eqnarray}
    H^{2,x(y)}_{\rm eff} = \sum_{\langle ij \rangle, \delta= 4}\frac{J_K^2}{\Delta}\langle \sigma_i^{x(y)}\sigma_j^{x(y)}\rangle S_i^{x(y)}S_j^{x(y)}
\end{eqnarray}
where $\delta=4$ denotes black bonds in  in Fig.~\ref{Fig:2}(f) that interact via the $\Gamma^4$ operators. We calculate $\langle \sigma_i^{x(y)}\sigma_j^{x(y)}\rangle = {\rm sgn}(K) 0.497$ and $\Delta = 0.199|K|$. There are no higher-spin interactions since the black bonds are not connected with each other.

Next, we derive the $z$ component of the induced interaction  between impurity moments, $J_{\rm{eff}}^{z}\equiv J_z$ which is long-ranged. We follow the same method of exact diagonalization using Majorana fermions described in the previous section. We consider a two-sublattice unit cell, which is a convenient choice for a $\pi$-flux square lattice. As can be seen in Fig.~\ref{Fig:5}(a) and (b),  the interaction between magnetic moments on the  same sublattice, $J_{\rm{eff}}^{z,\rm{AA}}$,  is ferromagnetic, while the interaction between opposite sublattices, $J_{\rm{eff}}^{z,\rm{AB}}$,  is antiferromagnetic. Consequently, the long-range interactions are not frustrated.

  We again  use Eq.~(\ref{eq:RKKYGraphene}) to fit our numerical results,  but now with ${\bf k_D}=(\pm \frac{\pi}{2a}, 0)$. The use of this formula, originally derived for graphene, is justified by the fact that the low-energy spectra of graphene and the $\pi$-flux square lattice model are essentially identical, differing only in the locations of the Dirac cones within the Brillouin zone,
 as illustrated  in Fig.\ref{Fig:6} (a).  Therefore, there are  no oscillations along the $x$-direction yet we still observe $1/R^3$ decay. We obtained the best fit for $C=3.4\times 10^{-2}$. 

  Combining the perturbative short-ranged and RKKY-type long-ranged contributions, we obtain the induced  coupling Hamiltonian,

 
\begin{equation}
    H_{\rm eff}=\sum_{ij}J_{z}({\bf R}_i-{\bf R}_j) S_i^zS_j^z + J_{xy}\sum_{\langle ij \rangle_ {\delta= 4}} (S_i^xS_j^x+S_i^yS_j^y) \label{eq:HeffSql}
\end{equation}
where $J_{xy}={\rm sgn}(K)J_K^2 \langle \sigma_i^x\sigma_j^x\rangle/\Delta \simeq 2.497J_K^2/K$. We note that even tough the $z$-component has long-ranged unfrustrated interactions, the overall scale is about two orders of magnitude smaller than the $J_{xy}$ coupling on the $\delta=4$ bond. Since both $J_z$ and $J_{xy}$ scale with $J_K/K$, their ratio does not depend on $J_K/K$ for $J_K/K \ll 1$. 

In order to determine the ground state of Eq.~\ref{eq:HeffSql}, we first consider its limiting cases. For $J_{xy}=0$, the long-range Ising interaction leads to an Ising antiferromagnetic (AFM) ground state. In the other limit, $J_{z}=0$, the ground state 
 consists of isolated dimers on the black bonds, forming a product state. 
 These dimers have the wave function $|0,0\rangle = (\uparrow \downarrow - \downarrow \uparrow)/\sqrt{2}$ for  $J_{xy}>0$ ($K>0$) or $|1,0\rangle = (\uparrow \downarrow + \downarrow \uparrow)/\sqrt{2}$ for $J_{xy}<0$ ($K<0$). In both cases, the ground state is non-degenerate with an excitation gap of $2|J_{xy}|$ to $|1,1\rangle$ and $|1,-1\rangle$ states. This state does not break any symmetry of  effective Hamiltonian (\ref{eq:HeffSql}), so  we refer to it as a dimerized quantum paramagnet (DQP). The competition between $J_z$ and $J_{xy}$ terms leads to distinct ground states, with the dominant interaction  determining the system's behavior. We found  that the ratio of $J_{xy}$ to the largest  interaction  $J_z^ {\rm max}$, corresponding to the  nearest-neighbor interaction, is about $ 44$, so most of the phase space is occupied by the DQP  state.    
 
While the ratio $J_{xy}$ to $J_z^ {\rm max}$ is fixed in the  derived model, it can be tuned via the flux-coupling term  $J_w$, as discussed in the next section.

\subsubsection{  Tuning the effective couplings  $J_{xy}$ and $J_z$ in Eq.(\ref{eq:HeffSql})}

Since the ratio $J_{xy}/J_{z}$ determines the ground state of  the effective model Eq.(\ref{eq:HeffSql}), it is useful to introduce an additional term that directly controls this ratio. This can be achieved by adding a flux-coupling term, $-J_w\sum_{p,q}(\mathcal{W}_p^{SqL}+\mathcal{W}_q^{SqL})$, to Eq.(\ref{eq:SqL}). While this term does not modify the Hamiltonian's eigenstates, it shifts the energy of different flux sectors. For $J_w>0$, 
it introduces an energy cost to the $\pi$-fluxes  with $\mathcal{W}_{p/q}^{SqL}=-1$, thus eventually driving a transition to the $0$-flux state.

We can see this directly by noting that the energy of the $\pi$-flux sector is given by
 $E_{\pi}(J_w)=E_{\pi}(0)+NJ_w$ and the energy of the $0$-flux sector by
$E_0(J_w)=E_0(0)-NJ_w$, where $E_{\pi}(0)$ and $E_0(0)$ 
are their respective energies at $J_w=0$, and $N$ is the number of plaquettes. The two energy levels cross at a critical value
$J_w^c$,  satisfying  $E_{\pi}(0)+NJ_w^c=E_{0}(0)-NJ_w^c$. Solving for $J_w^c$ gives $J_w^c=(E_0(0)-E_{\pi}(0))/2N$, which evaluates to $J_w^c=0.0738$. Beyond this point, we assume that the ground state favors the $0$-flux configuration, while recognizing that a more precise analysis may reveal intermediate states.
 
We then compute the vison gap energy as a function of $J_w$ in the two distinct regions:
$\Delta(J_w)=\Delta_{\pi}-2J_w$ for $J_w<J_w^c$ in the $\pi$-flux sector and $\Delta(J_w)=\Delta_{0}+2J_w$ for $J_w>J_w^c$ in the $0$-flux sector. $\Delta_{\pi}$ and $\Delta_{0}$ represent the energy differences associated with flipping a single flux in the $\pi$- and $0$-flux sectors at $J_w=0$, respectively. We calculate $\Delta_{\pi}$ and $\Delta_{0}$ by considering a uniform $\pi$ and $0$ flux configurations and creating single vison excitations for $J_w=0$. Our calculations gives $\Delta_0=-0.135|K|$ and $\Delta_\pi=0.199|K|$. Note that $\Delta_0<0$, since $0$-flux is an excited state for $J_w=0$. In addition, we calculate the nearest-neighbor spin-spin correlation as $\langle \sigma_i^{x(y)}\sigma_j^{x(y)}\rangle = {\rm sgn}(K) 0.396$ in the $0$-flux sector which is used next in determining the $J_{xy}$ exchange.

Fig.~\ref{Fig:6}(b) illustrates how the vison gap energy and the ratio of $J_{xy}$ to $J_z^{\rm max}$, where $J_z^{\rm max}=\rm max\{|J_z|\}$, vary with $J_w$. The vison gap energy decreases with increasing $J_w$ up to a critical value of $J_w=0.0738|K|$, where the system undergoes a transition in the ground-state flux sector from $\pi$-flux sector to a 0-flux sector. Beyond this critical point, the vison gap increase linearly with $J_w$. At the transition point, the ratio $J_{xy}/J_z^{\rm max}$ reaches its maximum value. 

Since $J_{xy}\sim K^2/\Delta (J_w)$, it is possible to tune $J_{xy}$ with $J_w$. However, the $J_w$ term does not affect the eigenstates and thus does not directly influence the effective RKKY-like $J_z$ interaction within a given flux sector. Instead, the change in the $J_z$  interaction occurs through the flux sector transition. Namely, in the $0$-flux sector of the square lattice, the perfectly nested Fermi surface with a van Hove singularity at the Fermi level  (see Fig.\ref{Fig:6}(a)) induces an oscillatory RKKY interaction, with a spatial period determined by the wave vector  ${\bf q}= (\pi/a,\pi/a)$. 
Fig.~\ref{Fig:5}(c) and (d) illustrate the behavior of the RKKY interaction in the 0-flux sector, emphasizing the distinct decay rates along different lattice directions as previously reported\cite{Glodzik}. For spins aligned along the diagonal direction (Fig.~\ref{Fig:5}(c)), $J_{\rm eff}^{\rm diag}$ does not display an oscillatory behavior, since ${\bf q}\cdot ({\bf r}_{i}-{\bf r}_{j}) = 2\pi n$ ($n$ is the number of sites between $i$ and $j$). In contrast, for spins aligned along the $x$-direction (Fig.~\ref{Fig:5}(d)), the interaction, $J_{\rm eff}^x$, shows oscillatory behavior governed by ${\bf q}\cdot ({\bf r}_{i}-{\bf r}_{j}) = \pi n$. Additionally, the strength of $J_z^{\rm max}$ in the $0$-flux sector is nearly twice as large as that in the $\pi$-flux sector, indicating a significant enhancement of the interaction due to the presence of a Fermi surface and a van Hove singularity.

\begin{table}[t]
\begin{tabular}{|c|c|c|c|}
\hline
Model &  \# MF flavors  &  II &  GS \\
\hline
Kitaev & 1 & Kitaev &QSL\\
\hline
Yao-Lee & 3&  Heisenberg & AFM\\
\hline
SqL& 2 & Ising + XY & DQP, $J_w \in[-2.81,0.79]$ \\
& & &AFM, $J_w \notin [-2.81,0.79]$ \\
\hline
\end{tabular}
\caption{\label{Table} 
 Summary of model properties, including the number of Majorana fermion flavors (\# MF), induced interactions (II), and ground states (GS): QSL for the Kitaev model; AFM for the Yao-Lee model; DQP for $J_w \in $ [-2.81,0.79]  and otherwise AFM state  for the SqL model.}
\end{table}

\begin{figure}[t]
\includegraphics[width=0.9\linewidth]{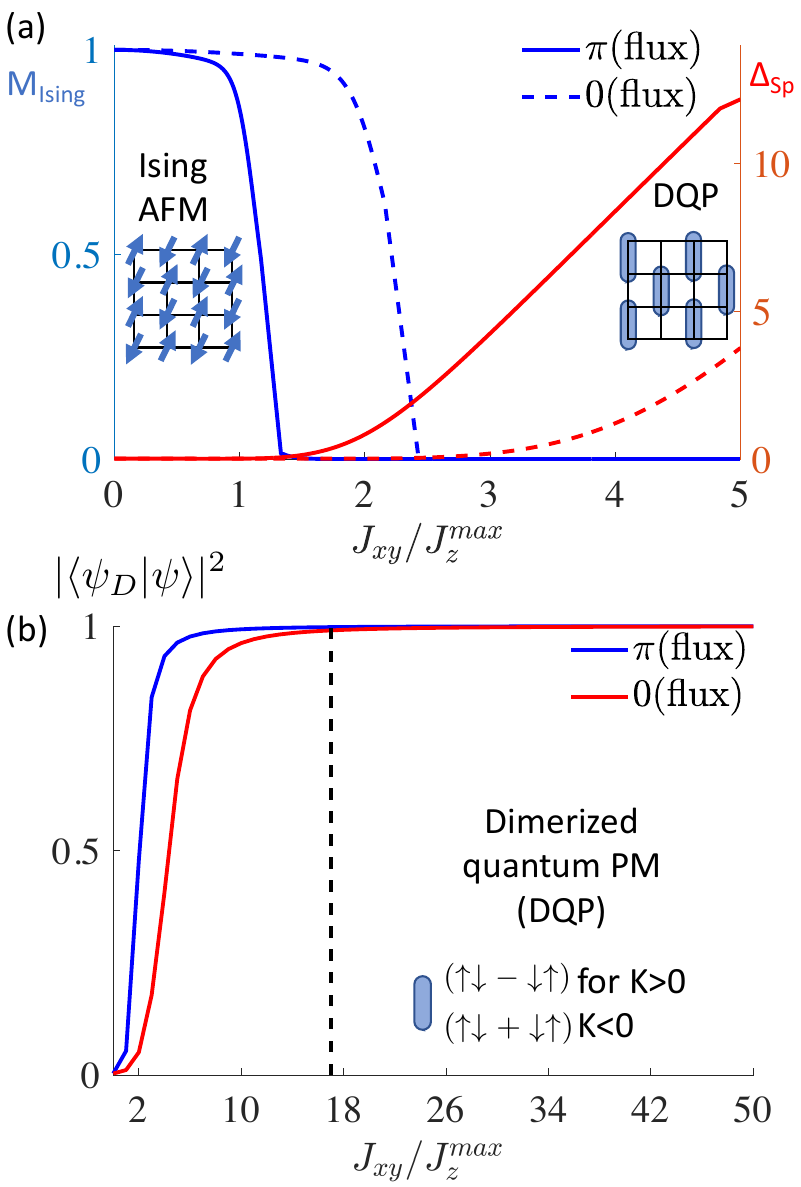}
    \caption{(a) Phase diagram of the induced Hamiltonian in the square lattice model  (Eq.~\ref{eq:HeffSql}) obtained by 16 site exact diagonalization. For $J_{xy}/J_z \gtrsim 1.5{(2.4)}$ in the $\pi$-flux(0-flux) sector, the Ising order parameter vanishes and a gap ($\Delta_{\rm Sp}$) opens between the ground state and the excited states as expected in a dimerized quantum paramagnet. (b) Overlap of the ground state wavefunction with an ideal dimerized quantum paramagnet, $|\psi_D\rangle$. For $J_{xy}/J_z=17$, the overlap $|\langle \psi_D | \psi \rangle|^2 \geq 0.99$ (black dotted line), indicating a nearly perfect overlap in both $\pi$-flux and 0-flux sectors.}
    \label{Fig:7}
\end{figure}
\subsubsection{ Ground state phase diagram of the tunable  model (\ref{eq:HeffSql}): ED  results on the 16-site cluster}

In Fig.~\ref{Fig:7}(a), we present exact diagonalization results for the effective Hamiltonian (\ref{eq:HeffSql}) on a 16-site cluster, considering both the $\pi$-flux and $0$-flux sectors. The long-range part of the $J_z$ term is retained up to the fourth nearest neighbor, as permitted by system size, and the results are shown as a function of $J_{xy}/J_{z}^{\rm max}$. In the $\pi$-flux sector, the ground state remains doubly degenerate for $J_{xy}/J_{z}^{\rm max}\lesssim 1.5$, while in the $0$-flux sector, this degeneracy persists up to 
$J_{xy}/J_{z}^{\rm max}\lesssim 2.4$.

We determine the AFM Ising order parameter $M_{\rm Ising} = \sum_i (M_i^A - M_i^B)/N$ where $M_i^{A(B)}$ is the magnetization in $A(B)$ sublattice by introducing a weak ($10^{-4}K$) staggered field that splits the degenerate states. For $J_{xy}/J_{z}^{max}\gtrsim  1.5$ in the $\pi$-flux  and $J_{xy}/J_{z}^{max}\gtrsim  2.4$ in the $0$-flux, the Ising order parameter vanishes and a gap between the ground state and first excited state ($\Delta_{\rm Sp}$) appears which implies the stabilization of the DQP phase.

In Fig.~\ref{Fig:7}(b), we present the overlap between the ground state wave function $|\psi \rangle$ and the dimerized wave function $|\psi_D \rangle$ computed with $J_z=0$. Even with $J_w=0$, the fixed value $J_{xy}/J_{z}=44$ gives  $|\langle \psi_D|\psi\rangle|^2 
\simeq 0.9997$, implying that the ground state of the induced Hamiltonian for the square lattice model lies in the DQP phase. As $J_w$ varies, the ratio $J_{xy}/J_{z}$ changes: it is approximately 1.5 at $J_w=-2.81|K|$ and 2.4 at $J_w=0.79|K|$. This suggests that within the range $-2.81 \lesssim J_w\lesssim0.79$, the ground state remains in the DQP phase. However, outside this range, the ground state transitions to the Ising antiferromagnetic phase. 

\section{Conclusion}
In this work, we investigated the effective exchange interactions among local magnetic moments mediated by fractionalized excitations in the Kitaev-type QSLs. By considering three models --   the Kitaev model, the Yao-Lee  model, and a $\pi$-flux square-lattice generalization (SqL) -- we explored the nature of induced interactions and their implications for magnetic order.  The summary of  key properties for each model, including the number of Majorana fermion flavors (\# MF) flavors, the nature of induced interactions (II), and the resulting ground states (GS)  is presented in Table \ref{Table}.

Our findings reveal that the commutation relation between spin operators and flux operators plays a decisive role in determining the interaction range. In the Kitaev model, where spin operators anticommute with flux operators, the induced interactions are short-ranged and can be treated perturbatively. These interactions result in the effective Hamiltonian for the local moments that retains the structure of the Kitaev model but with renormalized couplings, thereby ensuring the stability of the spin-liquid ground state. Including higher-order terms in the perturbative treatment extends the range of these interactions while preserving the exact solvability of the model.

In contrast, the YL model, characterized by the presence of three flavors of Majorana fermions and SU(2) spin symmetry, supports long-range, RKKY-type interactions mediated by itinerant Majorana fermions. These interactions stabilize an AFM ground state, partially gapping the Majorana fermion spectrum. The similarity between the induced interaction in the YL model and the RKKY interaction in graphene highlights the impact of the Dirac spectrum on the coupling range and oscillatory behavior.

The SqL model exhibits a rich interplay between short- and long-range interactions. The short-range interactions are highly anisotropic, favoring specific bond directions, while the long-range interactions between $z$-components of the magnetic moments are unfrustrated and decay with distance as $1/R^3$. The resulting ground state is a dimerized quantum paramagnet (DQP) in the $\pi$-flux sector. By introducing a term that shifts the flux-sector energetics, we demonstrated the tunability of the SqL model, leading to a  transition  from $\pi$-flux  into a 0-flux sector. This transition alters the ratio of short- to long-range couplings between impurity moments. Depending on the value of this tuning term $J_w$,  the local moments can have  two distinct ground states:  a DQP for $J_w \in [-2.81,0.79]$, and an AFM state for values of $J_w$ outside this range.

Our study illustrates the diverse range of magnetic orders and interaction mechanisms that can arise from coupling local moments to Kitaev-type spin liquids. It might also  be useful for understanding the interplay between fractionalized excitations and magnetic impurities \cite{Takahashi2025}, emphasizing the potential of Kitaev materials to host exotic magnetic orders mediated by Majorana fermions.

Future directions include exploring the competition between the Kondo effect and magnetic order at larger Kondo couplings and extending the analysis to anisotropic QSLs or systems with more complex geometries. These extensions could reveal additional phases and deepen our understanding of quantum magnetism in strongly correlated systems.

\section{Acknowledgements}
We thank Johannes Knolle and Shi Feng for fruitful discussions. O.E. acknowledges support from NSF Award No. DMR-2234352. M. A. Keskiner would like to gratefully acknowledge the financial support provided by T\"{U}B\.{I}TAK under the T\"{U}B\.{I}TAK B\.{I}DEB 2214-A Yurt Dı\c{s}ı Doktora Sırası Ara\c{s}tırma Burs Programı (T\"{U}B\.{I}TAK B\.{I}DEB 2214-A - International Research Fellowship Programme for PhD Students). M. A. Keskiner also acknowledges the hospitality of ASU during his visit.  M.O.O is partially supported by T\"{U}B\.{I}TAK. 1001 grant No. 122F346.
The work by N.B.P. was supported by the  National Science Foundation under Award No.\ DMR-1929311.  N.B.P. also
acknowledges the hospitality and partial support of the Technical University of Munich – Institute for Advanced Study and the support of the Alexander von Humboldt Foundation.

\appendix
\section{Higher order induced exchange interactions in Kitaev model} \label{App:A}
\begin{figure}[t]
    \includegraphics[width=1\linewidth]{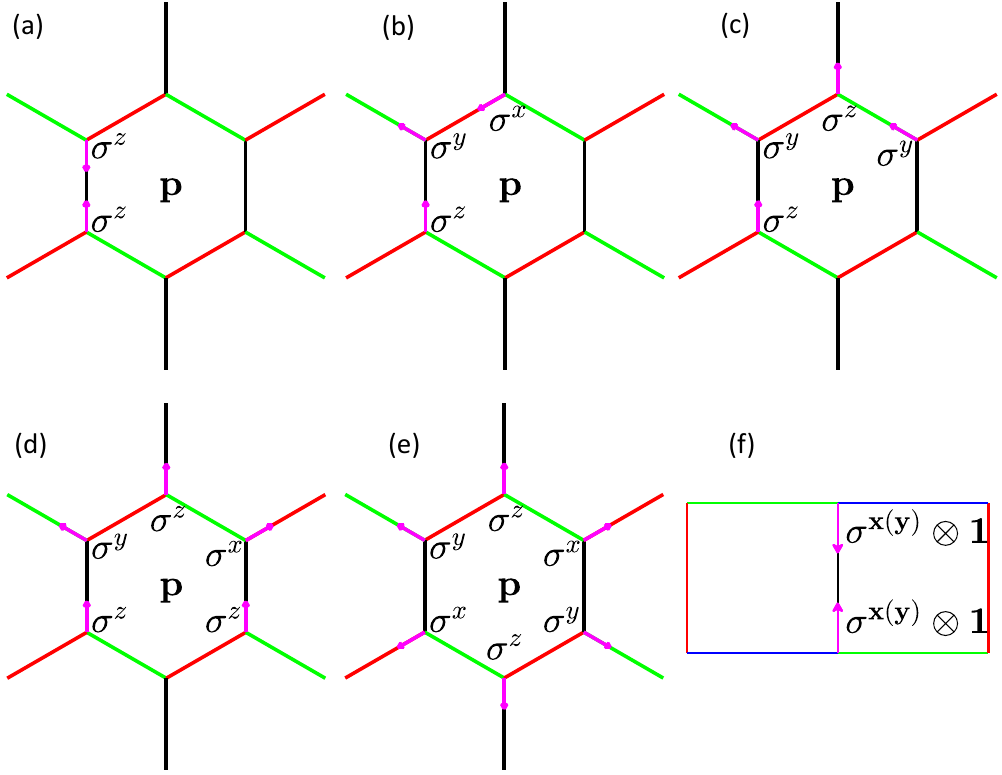}
    \caption{In Kitaev model, applying the $\sigma^{\alpha}$ operator to the ground state wave function creates two $\pi$-fluxes. To return the ground state flux sector, $\sigma$ operators can be placed on plaquette sites in distinct ways as depicted in (a)-(e). The ground state of the square lattice model lies in the $\pi$-flux sector. When the $\sigma_{x(y)}$ operator is applied, it flips the $u_{ij}$ defined on the black bond ($\Gamma_i^4\Gamma_j^4$ interaction), creating two 0-fluxes.  To restore the system to the ground state flux sector, the same operator must be applied to the other site linked by the black bond.}
    \label{Fig:8}
\end{figure}

In the Kitaev model, applying a spin operator ($\sigma^{\alpha}$) to a site creates two $\pi$-fluxes  through the adjacent plaquettes separated by the corresponding bond type $(\alpha=x,y,z)$. To restore the ground state 0-flux sector, subsequent spin operators can be applied in a systematic way based on the bond configuration. In configuration Fig.~\ref{Fig:3}(a), two nearest-neighbor $\sigma^z$ operators are applied on sites linked by a black bond, where the first creates two fluxes and the second flips them back, restoring the ground state. In  Fig.~\ref{Fig:8}(b) three spin operators $(\sigma^z, \sigma^y, \sigma^x)$ are applied sequentially: $\sigma^z$ generates fluxes through the black-bond plaquettes, $\sigma^y$ flips fluxes through the green-bond plaquettes (partially canceling the initial fluxes), and $\sigma^x$ restores the 0-flux sector by flipping the fluxes through the red-bond plaquettes. Similarly, configurations (c), (d), and (e) involve distinct sequences of spin operators acting on specific bonds  ($z$ on black, $y$ on green, and $x $ on red), with each spin operator sequentially creating, flipping, or annihilating fluxes. The arrangement of spin operators in these configurations also respects the $C_3$ symmetry of the system.

Considering three magnetic impurities that are placed on a plaquette, as depicted in Fig.~\ref{Fig:8}(b), the perturbation term becomes 
$V=J_K({S}_i^z\sigma_i^z+{S}_j^y\sigma_j^y+{S}_k^x\sigma_k^x)$, and the third-order correction yields the induced interaction between spins, given by
\begin{eqnarray}
  H_{eff}^{(3)}&=&\frac{3J_K^3}{2\Delta ^2}\langle\sigma_i^z\sigma_j^y\sigma_k^z\rangle S_i^zS_j^yS_k^x\\  &=&\frac{3J_K^3}{2\Delta^2}\langle ib_i^zc_iib_j^yc_jib_k^xc_k \rangle S_i^zS_j^yS_k^x\\ 
  &=&\frac{3J_K^3}{2\Delta^2}\langle -ic_ic_k\rangle S_i^zS_j^yS_k^x\\  
  &=&0 
\end{eqnarray}
where we fixed the gauge to $u_{ij}=1$. The expectation value of $\langle -ic_ic_k \rangle$
is zero due to time reversal symmetry. Next, we determine the induced interactions for the configurations shown in Fig.~\ref{Fig:3}(c),(d), and (e) as $ H_{\rm eff}^{(4)}=\frac{3J_K^4}{\Delta^3}\langle-ic_ic_l\rangle S_i^zS_j^yS_k^zS_l^y$ with $\langle-ic_ic_k\rangle =0.186$, $H_{\rm eff}^{(5)}=\frac{15J_K^5}{2\Delta^4}\langle-ic_ic_m\rangle S_i^zS_j^yS_k^zS_l^xS_m^z
 =0$, and $H_{\rm eff}^{(6)}=\frac{45J_K^6}{2 \Delta^5}\langle\mathcal{W}_p\rangle S_i^xS_j^yS_k^zS_l^xS_m^yS_n^z=\frac{45J_K^6}{2 \Delta^5} S_i^xS_j^yS_k^zS_l^xS_m^yS_n^z$, respectively.
 
 \section{The exact solvability of the  Hamiltonian given by Eq.~\ref{eq:effKitaev}} \label{App:B}
 The third term only includes the plaquette operator, and since $J_6>0$, it attempts to flip the fluxes. However, since $J_6\ll J_2$, the fluxes do not change, so we can ignore this term.
 Since $S^{\alpha}S^{\beta}=i\epsilon_{\alpha\beta\gamma}S^{\gamma}$, we can write the second term as $-S_i^{\alpha}S_j^{\alpha}S_j^{\gamma}S_k^{\gamma}S_k^{\beta}S_l^{\beta}$. In terms of Majorana fermions, it can be written as $-i(ib_i^{\alpha}b_j^{\alpha})(ib_j^{\gamma}b_k^{\gamma})(ib_k^{\beta}b_l^{\beta})c_ic_l=-iu_{ij}u_{jk}u_{kl}c_ic_l$. By setting $u$ to achieve the ground state and and taking the Fourier transform of the Majorana fermions, $c_i^{a,b}=\sqrt{\frac{2}{N}}\sum_{k\in HBZ} e^{i\vec{k}.\vec{r}}c_k^{a,b}+e^{-i\vec{k}.\vec{r}}{c_k^{a,b}}^{\dagger}$, where $N$ is the number of unitcells and $a$ and $b$ represent the even and odd sublattices, respectively, the Hamiltonian can be represented in $k$-space as follows
\begin{equation}
    H=\sum_{k\in HBZ}\begin{pmatrix}
       a_{k}^{\dagger}&b_{k}^{\dagger} 
    \end{pmatrix}
    \begin{pmatrix}
        0 & i(f_k+g_k)\\
        -i(f_k^{*}+g_k^{*}) & 0
    \end{pmatrix}\begin{pmatrix}
        a_{k}\\b_{k}
    \end{pmatrix}
    \end{equation}
where $a_{k}$ and $b_{k}$ denote  fermions the even and odd sublattices, respectively; $f_k=2J_2(1+e^{-ik_1}+e^{-ik_2})$ and $g_k=2J_4(e^{-i(k_1-k_2)}+e^{-i(k_1+k_2)}+e^{-i(k_2-k_1)})$ with $k_1={\bf k} \cdot {\bf b}_1$ and $k_1={\bf k}\cdot{\bf b}_2$ where ${\bf b}_1$ and ${\bf b}_2$ are the primitive lattice vectors.

\section{Mapping the Yao-Lee spin liquid with Kondo interaction to a graphene-like model via complex fermions} \label{App:C}
The z-component of the Kondo coupling, $J_K\sum_iS_i^zic_i^xc_i^y$,  selectively couples two out of the three flavors of Majorana fermion, namely $c^x$ and $c^y$, while leaving the third flavor, $c^z$, decoupled. This allows us to disregard $c^z$ and reformulate the problem in terms of a complex fermion representation. By defining a complex fermion as $f_i=(c_i^x+ic_i^y)/2$, the model can be mapped onto a graphene-like system. In this mapping, the electron and hole bands of the new model correspond to the spin-up and spin-down bands of graphene. In the ground-state flux sector, our Hamiltonian, when expressed in terms of complex fermions, can be written as the summation of the Hamiltonians for particles and holes: 
\begin{equation}
    H=H_p+H_h
\end{equation}
where 
\begin{equation}
    H_p=\sum_{\langle ij\rangle}K(if_i^{\dagger}f_j+h.c.)-J_K\sum_if_i^{\dagger}f_i
\end{equation}
represents the particle sector, and
\begin{equation}
    H_h=\sum_{\langle ij\rangle}K(if_if_j^{\dagger}+h.c.)+J_K\sum_if_if_i^{\dagger}
\end{equation}
corresponds to the hole sector. This decomposition allows us to interpret the system in analogy with a graphene-like band structure, where particle and hole excitations play roles similar to spin-up and spin-down bands in graphene.

\bibliography{references.bib}
\end{document}